\documentclass{article}
\usepackage{amsmath, amssymb, amsfonts}
\title{On a conformal Schwarzschild-de Sitter spacetime}  
\author{Hristu Culetu, \\Ovidius University, Dept. of Physics and Electronics \\ Bld. Mamaia 124, 900527 Constanta, Romania \footnote{electronic address: hculetu@yahoo.com}}

\begin{document}
\numberwithin{equation}{section}
\pagenumbering{arabic}
\maketitle
\newcommand{\fv}{\boldsymbol{f}}
\newcommand{\tv}{\boldsymbol{t}}
\newcommand{\gv}{\boldsymbol{g}}
\newcommand{\OV}{\boldsymbol{O}}
\newcommand{\wv}{\boldsymbol{w}}
\newcommand{\WV}{\boldsymbol{W}}
\newcommand{\NV}{\boldsymbol{N}}
\newcommand{\hv}{\boldsymbol{h}}
\newcommand{\yv}{\boldsymbol{y}}
\newcommand{\RE}{\textrm{Re}}
\newcommand{\IM}{\textrm{Im}}
\newcommand{\rot}{\textrm{rot}}
\newcommand{\dv}{\boldsymbol{d}}
\newcommand{\grad}{\textrm{grad}}
\newcommand{\Tr}{\textrm{Tr}}
\newcommand{\ua}{\uparrow}
\newcommand{\da}{\downarrow}
\newcommand{\ct}{\textrm{const}}
\newcommand{\xv}{\boldsymbol{x}}
\newcommand{\mv}{\boldsymbol{m}}
\newcommand{\rv}{\boldsymbol{r}}
\newcommand{\kv}{\boldsymbol{k}}
\newcommand{\VE}{\boldsymbol{V}}
\newcommand{\sv}{\boldsymbol{s}}
\newcommand{\RV}{\boldsymbol{R}}
\newcommand{\pv}{\boldsymbol{p}}
\newcommand{\PV}{\boldsymbol{P}}
\newcommand{\EV}{\boldsymbol{E}}
\newcommand{\DV}{\boldsymbol{D}}
\newcommand{\BV}{\boldsymbol{B}}
\newcommand{\HV}{\boldsymbol{H}}
\newcommand{\MV}{\boldsymbol{M}}
\newcommand{\be}{\begin{equation}}
\newcommand{\ee}{\end{equation}}
\newcommand{\ba}{\begin{eqnarray}}
\newcommand{\ea}{\end{eqnarray}}
\newcommand{\bq}{\begin{eqnarray*}}
\newcommand{\eq}{\end{eqnarray*}}
\newcommand{\pa}{\partial}
\newcommand{\f}{\frac}
\newcommand{\FV}{\boldsymbol{F}}
\newcommand{\ve}{\boldsymbol{v}}
\newcommand{\AV}{\boldsymbol{A}}
\newcommand{\jv}{\boldsymbol{j}}
\newcommand{\LV}{\boldsymbol{L}}
\newcommand{\SV}{\boldsymbol{S}}
\newcommand{\av}{\boldsymbol{a}}
\newcommand{\qv}{\boldsymbol{q}}
\newcommand{\QV}{\boldsymbol{Q}}
\newcommand{\ev}{\boldsymbol{e}}
\newcommand{\uv}{\boldsymbol{u}}
\newcommand{\KV}{\boldsymbol{K}}
\newcommand{\ro}{\boldsymbol{\rho}}
\newcommand{\si}{\boldsymbol{\sigma}}
\newcommand{\thv}{\boldsymbol{\theta}}
\newcommand{\bv}{\boldsymbol{b}}
\newcommand{\JV}{\boldsymbol{J}}
\newcommand{\nv}{\boldsymbol{n}}
\newcommand{\lv}{\boldsymbol{l}}
\newcommand{\om}{\boldsymbol{\omega}}
\newcommand{\Om}{\boldsymbol{\Omega}}
\newcommand{\Piv}{\boldsymbol{\Pi}}
\newcommand{\UV}{\boldsymbol{U}}
\newcommand{\iv}{\boldsymbol{i}}
\newcommand{\nuv}{\boldsymbol{\nu}}
\newcommand{\muv}{\boldsymbol{\mu}}
\newcommand{\lm}{\boldsymbol{\lambda}}
\newcommand{\Lm}{\boldsymbol{\Lambda}}
\newcommand{\opsi}{\overline{\psi}}
\renewcommand{\tan}{\textrm{tg}}
\renewcommand{\cot}{\textrm{ctg}}
\renewcommand{\sinh}{\textrm{sh}}
\renewcommand{\cosh}{\textrm{ch}}
\renewcommand{\tanh}{\textrm{th}}
\renewcommand{\coth}{\textrm{cth}}

\begin{abstract}
On the basis of the C-metric, we investigate the conformal Schwarzschild - deSitter spacetime and compute the source stress tensor and study its properties, including the energy conditions. Then we analyze its extremal version ($b^{2} = 27m^{2}$, where $b$ is the deS radius and $m$ is the source mass), when the metric is nonstatic. The weak-field version is investigated in several frames, and the metric becomes flat with the special choice $b = 1/a$, $a$ being the constant acceleration of the Schwarzschild-like mass or black hole. This form is Rindler's geometry in disguise and is also conformal to a de Sitter metric where the acceleration plays the role of the Hubble constant. In its time dependent version, one finds that the proper acceleration of a static observer is constant everywhere, in contrast with the standard Rindler case. The timelike geodesics along the z-direction are calculated and proves to be hyperbolae.
 \end{abstract}
 
 \section{Introduction}
 The well-known C-metric describes a pair of uniformly accelerated black holes (BHs) in the Minkowski spacetime and it belongs to a class of spaces with boost-rotation symmetries \cite{RM, PG, JP1, AG}. Their acceleration is rooted from conical singularities produced by a strut between the two BHs or two semi-infinite strings connecting them to infinity. The pair creation of BHs may be possible in a background with a cosmological constant $\Lambda$ as this supplies the negative potential energy \cite{PD, MR}.
 
 To find the physical interpretation of the $\Lambda \neq 0$ case, Podolsky and Griffiths \cite{PG} introduced a new coordinate system adapted to the motion of two uniformly accelerating test particles in de Sitter (deS) space. However, the curvature singularity at $r = 0$ is still present. A physical meaning of the C-metric with a negative $\Lambda$ was given by Podolsky \cite{JP1}. He showed that this exact solution of Einstein's field equations describes uniformly accelerated BHs in anti de Sitter (AdS) universe, using a convenient coordinate system. More recently Arnfinnsson and Gron \cite{AG} (see also \cite{HC1}) found a new source (a singular accelerating mass shell) of the C-metric, using the Israel junction conditions. They took advantage of the C-metric in spherical coordinates, previously used in \cite{GKP}. The shell consists of a perfect fluid that creates a jump of the extrinsic curvature when the shell is crossed. Their metric corresponds to a nonlinear combination of the Schwarzschild and Rindler spacetimes,  thus representing the geometry outside an accelerated point-particle or black hole.

Our purpose in this work is to analyze in detail the conformal Schwarzschild-de Sitter (S-deS) spacetime. In Sec.2 we introduce the corresponding line-element, having a conformal factor rooted from the C-metric. The proper acceleration of a static observer and the components of the source stress tensor and its energy conditions are calculated. Sec.3 is devoted to the extremal case $b^{2}= 27m^{2}$, when the spacetime has a ''double'' horizon at $r = 3m$ and it is time dependent since $g_{tt}>0$. The weak field limit $m \rightarrow 0$ is treated in Sec.4, with a special choice $b = 1/a$ between the deS radius and the constant acceleration $a$. In that case the metric beomes flat (time dependent conformal deS space, being Rindler's in disguise), to which the timelike radial geodesics are computed. The proper acceleration of a static observer is constant in the whole space, in contrast with the Rindler case. A summary of the work is presented in Sec.5.

 Throughout the paper we use geometrical units $G = c = 1$, unless otherwise specified.
	
 \section{Conformal Schwarzschild-de Sitter metric}
  Let us write down the C-metric in spherical coordinates, following \cite{AG, GKP} 
    \begin{equation}
  ds^{2} = \frac{1}{(1 + ar cos\theta)^{2}} \left(- Q dt^{2} + \frac{1}{Q} dr^{2} + \frac{r^{2}}{P}d\theta^{2} + Pr^{2}sin^{2}\theta d\phi^{2}\right),
 \label{2.1}
 \end{equation}
 where $Q = (1 - a^{2}r^{2})(1 - 2m/r), P = 1 + 2amcos\theta$, $a$ is a positive constant and $m$ is the BH (or a point particle) mass. According to the authors of \cite{AG} and \cite{GKP}, Eq. (2.1) can be viewed as a nonlinear combination of the Schwarzschild and Rindler geometries, representing the metric around an accelerating point particle or a BH. The function $P$ is related to a conical singularity along the symmetry axis where the sources of the acceleration $a$ (a cosmic string or a strut) are supposed to be located.
 
 One observe that the term $2amcos\theta$ in the expression of $P$ is very low compared to unity when reasonable values for $a$ and $m$ are used. We have, indeed, $am = amG/c^{4} = ma/(c^{4}/G)$, where $c^{4}/G \approx 10^{44} N$ is a maximal force. Therefore, we shell work in the approximation $P = 1$. 

  We propose a different combination of the Schwarzschild (S) and de Sitter (deS) factors from the expression of $Q$ and write $Q = 1 - 2m/r - r^{2}/b^{2}$, as for the well-known S-deS line-element \cite{RB, GP}. The positive constant $b$ plays the role of the deS radius of curvature, with $b = \sqrt{3/\Lambda}$, $\Lambda$ being the cosmological constant. We will, however, preserve the first order term $ar/c^{2}$ in the conformal factor. The above approximations yield the conformal S-deS metric
     \begin{equation}
  ds^{2} = \frac{1}{(1 + ar cos\theta)^{2}} \left[- (1 - \frac{2m}{r} - \frac{r^{2}}{b^{2}}) dt^{2} + \frac{dr^{2}}{1 - \frac{2m}{r} - \frac{r^{2}}{b^{2}}}  + r^{2} d \Omega^{2}\right],
 \label{2.2}
 \end{equation}
 where $d \Omega^{2}$ stands for the metric on the unit 2-sphere. We might consider (2.2) as a metric by itself (with no deficit angle) and forget that it is rooted from the C-metric. 

 Let us take now a static observer in the geometry (2.2) with the velocity vector field
 \begin{equation}
 u^{b} = \left(\frac{1 + ar cos\theta}{\sqrt{1 - \frac{2m}{r} - \frac{r^{2}}{b^{2}}}}, 0, 0, 0\right), ~~~u^{a}u_{a} = -1,
 \label{2.3}
 \end{equation} 
 where $b, a$ labels $(t, r, \theta, \phi)$. The acceleration 4-vector $a^{b} = u^{a}\nabla_{a}u^{b}$ has the nonzero components
 \begin{equation}
 a^{r} = \left(\frac{m}{r^{2}} - \frac{r}{b^{2}} + \frac{3am~cos\theta}{r} - acos\theta \right)(1 + ar cos\theta),~~~ra^{\theta} = asin\theta (1 + ar cos\theta)
 \label{2.4}
 \end{equation} 
 with the proper acceleration
  \begin{equation}
	\begin{split}
  A \equiv \sqrt{a^{b}a_{b}} \\ = \sqrt{\frac{(\frac{m}{r^{2}} - \frac{r}{b^{2}} + \frac{3am~cos\theta}{r} - acos\theta )^{2} + (1 - \frac{2m}{r}-\frac{r^{2}}{b^{2}})a^{2}sin^{2}\theta}{1 - \frac{2m}{r} - \frac{r^{2}}{b^{2}}}}
 \label{2.5}
\end{split}
 \end{equation} 
We are dealing with an accelerating Schwarzschild-like particle or a black hole (BH), embedded in a deS universe. It is worth mentioning that, provided $P = 1$ in (2.2), we have no a conical singularity (no deficit angle) and so there is no need for a cosmic string or strut. The particle (or BH) might have an ordinary mass, not mandatory an astronomical one (same is valid for the radius $b$). It could be accelerated using obvious means.  

 Our independent parameters are the particle mass $m$, the acceleration $a$ and the deS radius $b$. As we see from (2.4), the radial acceleration depends on the radial and angular coordinates. It is worth noting that in the limit $b \rightarrow \infty$, $a^{r}$ and $A$ become the corresponding expressions from \cite{HC1}. In the expression of $a^{r}$ from (2.4), the first two terms within the first paranthesis give opposite contributions and at $r_{0} = (mb^{2})^{1/3}$ they cancel each other, the deS contribution dominating for $r > r_{0}$. Moreover, if we also consider the weak field limit $m \rightarrow 0$, the geometry (2.2) becomes conformally flat (anti de Sitter \cite{HC1}, with $\Lambda = -3a^{2}$) and $A = a$. That means the interpretation of the constant $a$ as an acceleration is justified.

 The horizons of the S-deS spacetime (2.2) are well studied (see, for example, \cite{JP2}). We only give here a short description. One denotes $f(r) \equiv 1 - 2m/r - r^{2}/b^{2}$ and look for its extremum. From the first derivative $f'(r) = df/dr$ 
 \begin{equation}
f'(r) = \frac{2m}{r^{2}} - \frac{2r}{b^{2}} = 0
 \label{2.6}
 \end{equation} 
one obtains only one real root $r_{0} = (mb^{2})^{1/3}$, whence $f(r_{0}) = 1 - 3m/r_{0}$. We distinguish three cases:

i) $r_{0}>3m$ (or $b^{2}>27m^{2}$). We get $f(r_{0}) >0$, such that one obtains two horizons: the BH horizon $r_{h}<r_{0}$ and the cosmological horizon $r_{c}>r_{0}$, $f(r)$ being positive between the two horizons (the 3rd root of $f(r) = 0$ is negative). 

The corresponding expressions of $r_{h}$ and $r_{c}$ are given by \cite{JP2}
 \begin{equation}
	 r_{h} = \frac{2b}{\sqrt{3}} cos(\frac{\alpha}{3} + \frac{4\pi}{3}), ~~~with~~~ cos\alpha = - \frac{3\sqrt{3}m}{b},~~~ 2m< r_{h} <3m	
 \label{2.7}
 \end{equation}
and
\begin{equation}
	 r_{c} = \frac{2b}{\sqrt{3}} cos\frac{\alpha}{3}, ~~~with~~~ r_{c}>3m.
 \label{2.8}
 \end{equation}

 ii) $r_{0}<3m$ (or $b^{2}<27m^{2}$), with $f(r_{0}) <0$. We have no horizons in this region.

iii) $r_{0} = 3m$ (or $b^{2} = 27m^{2}$). $f(r)$ can now be written as
\begin{equation}
f(r) = - \frac{1}{27m^{2}r} (r - 3m)^{2} (r + 6m),
 \label{2.9}
 \end{equation}
with a double root at $r = 3m$ ($r_{h}$ is increasing monotonically and $r_{c}$  is decreasing to the common value $r_{H} = 3m$, with $cos\alpha = -1$), when the two horizons coincide: $r_{b} = r_{c} = r_{0} = 3m$. This is the extremal (degenerate) situation \cite{JP2}, when $f(r) \leq 0$ for any positive $r$ and the metric (2.2) becomes nonstatic.

From the proper acceleration (2.5) we can find the surface gravity on each horizon (with $b^{2}>27m^{2}$, when $f(r) \geq 0, r\in[r_{b}, r_{c}]$)
 \begin{equation}
   \kappa = \sqrt{a^{b}a_{b}}~ \sqrt{-g_{tt}}|_{r = r_{H}} = |(1 - \frac{3m}{r_{H}})(\frac{1}{r_{H}} + acos\theta )|\frac{1}{|1 + ar_{H}cos\theta|} = \frac{1}{r_{H}}|(1 - \frac{3m}{r_{H}})|
 \label{2.10}
 \end{equation} 
where $r_{H}$ is one of the two horizons. In addition, $\kappa$ acquires a constant value (no dependence on $r$ or $\theta$), as it should be. Therefore, the discussion concerning the horizons resembles the $S-deS$ situation, without the conformal factor in (2.2). In addition, one notices that, when $r << b$ (or b tends to infinity), the geometry (2.2) is conformal to Schwarzschild's (see Ref.[8]), $r_{H} = 2m$ and $\kappa$ becomes $1/4m$, as expected. Moreover, $\kappa$ vanishes in the extremal situation $r_{0} = r_{H} = 3m$. That is similar with the extremal Reissner-Nordstrom geometry, where the surface gravity is also zero when $m = q$, $q$ being the BH (or particle) charge. 

As far as the scalar curvature is concerned, one obtains from (2.2) that
   \begin{equation}   
  R^{a}_{~a} = -12\left(a^{2} - \frac{1}{b^{2}} +  \frac{amcos\theta}{r^{2}} - \frac{ma^{2}cos^{2}\theta}{r} \right),  
\label{2.11}
\end{equation} 
which is divergent at $r = 0$.

 We look now for the sources of the spacetime (2.2), namely the stress tensor to lie on the r.h.s. of Einsteins' equations $G_{ab} = 8\pi T_{ab}$ in order that (2.2) to be an exact solution. One finds that
   \begin{equation} 
   \begin{split} 
  T^{t}_{~t} = -\rho = \frac{3}{8\pi}\left(a^{2} - \frac{1}{b^{2}} + \frac{2am}{r^{2}}cos\theta \right) ,~~~ T^{r}_{~r} = p_{r} = - \rho,\\ T^{\theta}_{~\theta} = T^{\phi}_{~\phi} = p_{\theta} = p_{\phi} =  \frac{3}{8\pi} \left(a^{2} - \frac{1}{b^{2}} - \frac{2ma^{2}}{r}cos^{2}\theta \right),  
\label{2.12}
\end{split}
\end{equation} 
where $\rho, p_{r}, p_{\theta}, p_{\phi}$ are respectively, the energy density, the radial pressure and the transversal pressures. It is worth noting that, in the weak field limit ($m \rightarrow 0$) the acceleration $a$ and the deS radius $b$ play opposed roles. In addition, the stress tensor is of $\Lambda$-type, $T_{ab} \propto g_{ab}$, the sign of $\bar{\Lambda} \equiv -3(a^{2} - 1/b^{2})$ depending on the relative values of $a$ and $1/b$. When $b = 1/a$, $\bar{\Lambda}$ and the scalar curvature are vanishing and the metric (2.2) becomes flat (with, of course, $m \rightarrow 0$). From (2.12) one sees that the energy conditions are not always satisfied. For instance, the sign of $\rho$ depends not only on the relative relation between $a$ and $b$ but also on the sign of $cos\theta$. Same is valid for the pressures.

 	The stress tensor (2.12) depends on too many independent parameters: $m, b$ and $a$, and so we need some relations between them for to investigate the energy conditions. 
	We distinguish three situations:
	
	1) $b = 1/a$
	
	That choice gives us
	\begin{equation} 
  \rho = - \frac{3am}{4\pi r^{2}}cos\theta = -p_{r},~~~ p_{\theta} = p_{\phi} = -\frac{3ma^{2}}{4\pi r}cos^{2}\theta,  
\label{2.13}
\end{equation} 
One observes that $\rho$ and $p_{r}$ switch their sign at $\theta = \pi/2$ but the transversal pressures remain always negative. For $\theta \in (0,\pi/2),~\cos\theta >0$ and the energy density $\rho$ is negative. In that case the energy conditions are not obeyed. When $\cos\theta <0$, $\rho$ becomes positive. To check the weak energy condition ($WEC$, $\rho \geq 0, \rho + p_{r} \geq 0, \rho + p_{\theta} \geq 0$ ), we evaluate
	\begin{equation} 
  \rho + p_{\theta} = \frac{3am|cos\theta|}{4\pi r^{2}} (1 - ar|cos\theta|), 
\label{2.14}
\end{equation} 
which is non negative for $r\leq 1/(a|cos\theta|)$. This condition may be satisfied in reasonable situations, when $1/a$ is very large. Not only the WEC is obeyed but also the null energy condition ($NEC$) and the dominant energy condition ($DEC$), because $\rho \geq |p_{r}|$ and $\rho \geq |p_{\theta}|$ are valid, too. Only the strong energy condition ($SEC$) is not satisfied, since $ \rho + p_{r} + 2p_{\theta}$ is negative.\\

   2) $a>>1/b$
	
	This case yields
		\begin{equation} 
  \rho \approx - \frac{3a^{2}}{8\pi} \left(1 + \frac{2mcos\theta}{ar^{2}}\right) = -p_{r},~~~ p_{\theta} = p_{\phi} \approx \frac{3a^{2}}{8\pi} \left(1 -  \frac{2mcos^{2}\theta}{r}.\right)
\label{2.15}
\end{equation} 
The above expressions for the components of the stress tensor have been already investigated in \cite{HC1}, Sec.2 and the process will not be repeated here.\\
 
    3) $a<<1/b$
		
	This case gives us 
			\begin{equation} 
		\rho \approx \frac{3}{8\pi b^{2}} - \frac{3am cos\theta}{4\pi r^{2}} = - p_{r},~~~p_{\theta} \approx -\frac{3}{8\pi b^{2}} - \frac{3a^{2}m cos^{2}\theta}{4\pi r}.
 		\label{2.16}
\end{equation} 
We firstly notice that the transversal pressures are always negative. Moreover, when $cos\theta \leq 0$, $\rho$ is positive everywhere. We get also $\rho + p_{\theta} \geq 0$ and $\rho \geq |p_{\theta}|$ if the condition $r\leq 1/(a|cos\theta|)$ is fulfilled (this is the same relation as that from the case $b = 1/a$). Therefore, all energy conditions are satisfied, excepting the SEC.

To be positive even for $cos\theta \geq 0$, the relation $r>b\sqrt{2amcos\theta}$ ought to be satisfied. But we work with $f(r)\geq 0$, such that $r \in [r_{h}, r_{c}]$, that leads us to $r<2b/\sqrt{3}$ (see Eqs.2.7 and 2.8). Keeping in mind that under the square root the very tiny ratio $ma/(c^{4}/G)$ appears ($c^{4}/G$ being the maximal force), we may say that the above inequality is generally obeyed. However, $\rho + p_{\theta} \leq 0$ always so the energy conditions are not satisfied.

\section{Extremal S-deS case}
We noticed before that the condition $b^{2} = 27m^{2}$ leads to the degenerate situation when the metric coefficient $f(r)$ has a double root at $r = 3m$, being otherwise negative for any positive $r$. Therefore, the radial coordinate becomes timelike and the temporal coordinate becomes spacelike (see \cite{HC2, DLC} for the Schwarzschild case). So, we interchange $t \rightarrow r$ and the metric (2.2) acquires the form
  \begin{equation}
  ds^{2} = \frac{1}{(1 + at cos\theta)^{2}} \left[- \frac{dt^{2}}{\frac{t^{2}}{27m^{2}} + \frac{2m}{t} - 1} + \left( \frac{t^{2}}{27m^{2}} + \frac{2m}{t} - 1\right) dr^{2}  + t^{2} d \Omega^{2}\right] ,
 \label{3.1}
 \end{equation}
with $t>0$ and non-negative metric coefficients that vanish at $t = 3m$. The line-element (3.1) is nonstatic and singular when $t \rightarrow 0$ (the same singularity as $r \rightarrow 0$ for the static version). Moreover, for the geometry (3.1) there is no a timelike Killing vector because the metric is time dependent.

If we take a ''static'' observer (laying at $r, \theta, \phi$ = const.)  with the velocity vector field
 \begin{equation}
 u^{b} = \left((1 + at cos\theta) \sqrt{\frac{t^{2}}{27m^{2}} + \frac{2m}{t} - 1}, 0, 0, 0\right), ~~~u^{a}u_{a} = -1,
 \label{3.2}
 \end{equation} 
one finds that $a^{r} = 0$ (since (3.1) is not r-dependent) and $ta^{\theta} = a sin\theta (1 + at cos\theta)$. For the proper acceleration, one obtains $A = asin\theta$. It is time independent, nonegative but not constant. That is a consequence of the fact that $a^{\theta}$ is the unique nonzero component of the 4-acceleration. An evaluation of the Ricci scalar gives us
   \begin{equation}   
  R^{a}_{~a} = -12a^{2} + \frac{4}{9m^{2}} + 12a^{2} \left(\frac{m}{t}cos^{2}\theta - \frac{m}{at^{2}}cos\theta \right),  
\label{3.3}
\end{equation} 
where we see the opposite contributions of the acceleration and the source mass.

 Let us study now the source stress tensor leading to the spacetime (3.1). One finds that
   \begin{equation} 
   \begin{split} 
  T^{t}_{~t} = -\rho = \frac{3}{8\pi}\left(a^{2} - \frac{1}{27m^{2}} + \frac{2am}{t^{2}}cos\theta \right) ,~~~ T^{r}_{~r} = p_{r} = - \rho,\\ T^{\theta}_{~\theta} = T^{\phi}_{~\phi} = p_{\theta} = p_{\phi} =  \frac{3}{8\pi} \left(a^{2} - \frac{1}{27m^{2}} - \frac{2ma^{2}}{t}cos^{2}\theta \right),  
\label{3.4}
\end{split}
\end{equation} 
We observe that the components of the above stress tensor have the same mathematical form as those from (2.2), with $t$ instead of $r$ and the time $t$ taking any non-negative value (with $-g_{tt}>0$ for any $t$). For example, when $t \rightarrow \infty$, the energy density and pressures in (3.4) no longer depend on the angular variable and the stress tensor is of $\Lambda$-type, with an effective $\Lambda_{eff} = -3a^{2} + 1/9m^{2}$. The same $\Lambda$-type form appears when we consider $a = 0$. In that case, $\Lambda_{eff} = 1/9m^{2}$. 

\section{Weak field limit}
Let us investigate now the properties of the $S-deS$ geometry in the weak field approximation ($m \rightarrow 0$) and with the special choice $b = 1/a$
      \begin{equation}
  ds^{2} = \frac{1}{(1 + ar cos\theta)^{2}} \left[- (1  - a^{2}r^{2}) dt^{2} + \frac{dr^{2}}{1 - a^{2}r^{2}}  + r^{2} d \Omega^{2}\right],~~~r<\frac{1}{a}
 \label{4.1}
 \end{equation}
that is conformal to the deS metric. It is also flat, as we have noticed at the end of the section 2 (see also \cite{AG}).

It is well-known that the static deS metric may be written in comoving coordinates, a version more appropriate in cosmology. The coordinate transformation is \cite{RT}
	 \begin{equation}
	\bar{r} = \frac{re^{-at}}{\sqrt{1 - a^{2}r^{2}}},~~~\bar{t} = t + \frac{1}{2a}ln(1 - a^{2}r^{2}),
 \label{4.2}
 \end{equation} 
and its inverse 
	 \begin{equation}
	r = \bar{r}e^{a\bar{t}},~~~t = \bar{t} - \frac{1}{2a} ln(1 - a^{2}\bar{r}^{2}e^{2a\bar{t}}).
 \label{4.3}
 \end{equation} 
In the barred coordinates, the geometry appears as
	 \begin{equation}
	ds^{2} = \frac{1}{(1 + a\bar{r}e^{a\bar{t}} cos\theta)^{2}}[-d\bar{t}^{2} + e^{2a\bar{t}} (d\bar{r}^{2} + \bar{r}^{2} d\Omega^{2})].
 \label{4.4}
 \end{equation} 
A static observer ($\bar{r}, \theta, \phi = const.$) with the velocity vector
	 \begin{equation} 
	 u^{b} = (1 + a\bar{r}e^{a\bar{t}} cos\theta, 0, 0, 0)
 \label{4.5}
 \end{equation} 
gives the 4-acceleration with the following nonzero components
 \begin{equation}
 a^{\bar{r}} = -a(1 + a\bar{r}e^{a\bar{t}} cos\theta) e^{-a\bar{t}} cos\theta,~~~\bar{r}a^{\theta} = a(1 + a\bar{r}e^{a\bar{t}} cos\theta) e^{-a\bar{t}} sin\theta
\label{4.6}
 \end{equation} 
As far as the proper acceleration is concerned, we get from (4.6) that $A = a$. Moreover, the scalar expansion is also constant, with $\Theta \equiv \nabla_{b}u^{b} = 3a$. It is nonzero due to the time dependence of the spacetime (4.4). The interesting property that $A$ is constant everywhere is a different feature from the Rindler metric in static coordinates, where the invariant acceleration is location dependent. That means any point of a spatially extended accelerating platform undergoes the same proper acceleration.

 Mathematically, $A$ is constant because the conformal factor from (4.4) simplifies (it appears symmetrically in the expressions of the components of acceleration from (4.6)). Same is valid for the exponential time dependent factors. The physical reason, in our view, behind the constancy of $A$ resides in some type of compensation between the time dependence of the line element (4.4), on the one hand, and the $r, \theta$ - dependence, on the other hand, through the conformal factor. In addition, the conformal factor has no influence on the value of the (constant) scalar expansion, which equals its deS value (the acceleration $a$ playing the role of the Hubble constant). 

It is useful to see what properties has the metric (4.4) in Cartesian coordinates $(\bar{t}, x, y, z)$
	 \begin{equation}
	ds^{2} = \frac{1}{(1 + aze^{a\bar{t}})^{2}}[-d\bar{t}^{2} + e^{2a\bar{t}} (dx^{2} + dy^{2} + dz^{2})].
 \label{4.7}
 \end{equation} 
An observer with the velocity vector
	 \begin{equation}
	 u^{b} = (1 + aze^{a\bar{t}}, 0, 0, 0)
 \label{4.8}
 \end{equation} 
will have a single nonzero component of the 4-acceleration: $a^{z} = -ae^{-a\bar{t}} (1 + aze^{a\bar{t}})$, the accelerating observer moving on the z-direction. We have again $A = a$, a constant proper acceleration. 

Our next task is finding the timelike geodesics in the time dependent form of the (flat) conformal deS geometry. Before that, we make an extra coordinate transformation, using the conformal time $\eta$ as the timelike variable
 	 \begin{equation}
	d\bar{t} = e^{a\bar{t}} d\eta ,~~~\eta = -\frac{1}{a} e^{-a\bar{t}}
 \label{4.9}
 \end{equation} 
The metric (4.7) becomes now
	 \begin{equation}
	ds^{2} = \frac{1}{a^{2} (z - \eta)^{2}} (-d\eta^{2} + dx^{2} + dy^{2} + dz^{2}).
 \label{4.10}
 \end{equation} 
With the velocity field $ u^{b} = [a(z-\eta), 0, 0, 0]$, one finds $a^{b} = [0, 0, 0, -a^{2} (z - \eta)]$ which leads again to $A = a$. However, the scalar expansion changes sign compared to the previous situations: $\Theta = -3a$, a possible consequence of the negative conformal time $\eta$.

We consider one is easier to take $T = -\eta >0$ and then to pass to the null coordinates $u = T+z,~v = T-z$. In terms of $u$ and $v$ (4.10) looks like 
	 \begin{equation}
	ds^{2} = \frac{1}{a^{2} u^{2}} (du dv + dx^{2} + dy^{2}).
 \label{4.11}
 \end{equation} 
 We will consider only the geodesics corresponding to $x = const.,~y = const.$. The starting Lagrangean appears as
	 \begin{equation}
	L = \frac{1}{a^{2} u^{2}} \dot{u} \dot{v},
 \label{4.12}
 \end{equation} 
where $\dot{u} = du/d\tau,~\dot{v} = dv/d\tau$, $\tau$ being the proper time. From (4.11) we have also $\dot{u} \dot{v} = a^{2} u^{2}$. The Euler-Lagrange equations give us the solutions
	 \begin{equation}
	v(\tau) = \tau,~~~u(\tau) = - \frac{1}{a^{2}\tau + k}
 \label{4.13}
 \end{equation} 
 with appropriate initial conditions ($k$ is a constant of integration). Once we get rid of $\tau$ and pass to the $T,z$ variables, one obtains 
	 \begin{equation}
	\left(z - \frac{k}{2a^{2}}\right)^{2} - \left(T + \frac{k}{2a^{2}}\right)^{2} = \frac{1}{a^{2}}.
 \label{4.14}
 \end{equation} 
 We consider $k = -2a$, a choice that will be justified later. Therefore, (4.13) appears as
	 \begin{equation}
	\left(z + \frac{1}{a}\right)^{2} - \left(T - \frac{1}{a}\right)^{2} = \frac{1}{a^{2}},
 \label{4.15}
 \end{equation} 
which represents two hyperbolae. At the moment $T = 1/a$ (or $\bar{t} = 0$), we get $z = 0$ or $z = -2/a$, corresponding to the two hyperbolae. The 2nd value of $z$ is not admissible because we must have $|-2/a| < \bar{r} < re^{-a\bar{t}} < r < 1/a$. With the previous boundary conditions, (4.15) yields
	 \begin{equation}
	z(T) = \sqrt{ \left(T - \frac{1}{a}\right)^{2} + \frac{1}{a^{2}}} - \frac{1}{a}.
 \label{4.16}
 \end{equation} 
One observes from (4.16) that we always have $z < 1/a$. That was the reason why we have chosen previously $k = -2a$. The curve $z(T)$ from (4.16) has its peak at $T = 1/a,~z = 0$ and asymptote $z = T - 2/a$ at infinity.

\section{Conclusions}
  A conformal S-deS geometry (an accelerating Schwarzschild-like mass or a BH embedded in a deS space) is examined in this work. The properties of the spacetime are investigated, in terms of the independent parameters $a, m$ and $b$. Even though the metric seems to have a complicate form, the source stress tensor looks simple and corresponds to an anisotropic fluid with different radial and transversal pressures. The condition $b^{2} = 27m^{2}$ leads to the extremal case when the two horizons coincide at $r = 3m$ and the spacetime becomes time dependent because the metric coefficient $f(r)$ is negative and the variable $r$ and $t$ exchange their roles.        
				
	The weak field limit $m \rightarrow 0$ with the special choice $b = 1/a$ is also studied. Even though the geometry is conformal to the deS one, it is surprisingly flat. The metric is the standard Rindler metric in disguise, having axial symmetry but written in spherical coordinates. It has a deS horizon at $r = 1/a$. We also studied the energy conditions corresponding to the stress tensor (2.12) for several values of the parameters.
	
The proper acceleration $A$ of a static observer ($r, \theta, \phi$-const.) equals $a$ at the origin of coordinates. We further use the Tolman coordinate transformation to write down the time dependent form of the conformal deS spacetime. In this situation the proper acceleration is $A = a$, irrespective of the location of the static observer in the accelerated system, in contrast with the standard Rindler frame. In addition, the scalar expansion of the same observer is also constant, $\Theta = 3a$. Going to the Cartesian coordinates and then changing the time variable to the conformal time $\eta$, one computes the timelike geodesics. They are hyperbolae with $z = T - 2/a$ as asymptote.\\

\textbf{Acknowledgements}

I am grateful to the anonymous referees for useful suggestions and comments which considerably improved the quality of the manuscript.

\end{document}